\documentclass[sigconf,screen=true,nonacm=true]{acmart}
\AtBeginDocument{%
  \providecommand\BibTeX{{%
    \normalfont B\kern-0.5em{\scshape i\kern-0.25em b}\kern-0.8em\TeX}}}

\usepackage{multirow}
\usepackage{amsmath}
\DeclareMathOperator*{\argmax}{arg\,max}

\usepackage{amsfonts}
\usepackage{algpseudocode}
\usepackage{algorithm}
\algnewcommand\algorithmicforeach{\textbf{for each}}
\algdef{S}[FOR]{ForEachpara}[1]{\algorithmicforeach\ #1\ \algorithmicdo \textbf{ in parallel}}
\algdef{S}[FOR]{ForEach}[1]{\algorithmicforeach\ #1\ \algorithmicdo}
\usepackage{xspace}
\newcommand{\FL}{{FL}\xspace}
\newcommand{\ML}{{ML}\xspace}
\newcommand{\scdg}{{SCDG}\xspace}
\newcommand{\scdgs}{{SCDGs}\xspace}
\newcommand{\lstm}{{LSTM}\xspace}
\newcommand{\gtwo}{{CDFS}\xspace}
\newcommand{\gone}{{CBFS}\xspace}
\newcommand{\gthree}{{BFS}\xspace}
\newcommand{\dsone}{{Dataset-1}\xspace}
\newcommand{\dstwo}{{Dataset-2}\xspace}

\DeclareMathOperator*{\x}{\mathbf{x}}

\DeclareMathOperator*{\Gr}{\mathbf{G}}
\DeclareMathOperator*{\Label}{\mathbf{L}}
\DeclareMathOperator*{\Gbar}{\mathbf{\widehat{G}}}

\DeclareMathOperator*{\EG}{\mathbf{E}(\mathbf{G})}
\DeclareMathOperator*{\Dx}{\mathbf{D}(\mathbf{x})}

\usepackage{tikz}
\usetikzlibrary{shapes.geometric, arrows}
\tikzstyle{startstop} = [rectangle, rounded corners, minimum width=1cm, minimum height=1cm,text centered, draw=black, fill=red!30]
\tikzstyle{io} = [trapezium, trapezium left angle=80, trapezium right angle=100, minimum width=1cm, minimum height=1cm, text centered, draw=black, fill=blue!30]
\tikzstyle{process} = [rectangle, minimum width=1cm, minimum height=1cm, text centered,text width=1.75cm, draw=black, fill=orange!30]
\tikzstyle{classifier} = [rectangle, minimum width=1cm, minimum height=3cm, text centered,text width=1cm, draw=black, fill=orange!30]
\tikzstyle{container} = [rectangle,minimum width=1.5cm, minimum height=3cm, draw=black, fill=orange!30]
\tikzstyle{classifier90} = [rectangle,dashed,rotate=90, minimum width=2.8cm, minimum height=0.4cm, text centered,text width=2cm, draw=black, fill=green!30]
\tikzstyle{decision} = [diamond, minimum width=3cm, minimum height=1cm, text centered, draw=black, fill=green!30]
\tikzstyle{arrow} = [thick,->,>=stealth]
\tikzstyle{arrow0} = [thick,<->,>=stealth]

\tikzstyle{note} = [text centered, text width=1cm,fill=yellow!10]

\tikzstyle{server} = [rectangle, rounded corners, minimum width=2cm, minimum height=1cm,text centered, text width=3cm, draw=black, fill=red!30]
\tikzstyle{client0} = [rectangle, rounded corners, minimum width=2cm, minimum height=1cm,text centered, text width=2cm, draw=black, fill=blue!30]
\tikzstyle{client1} = [rectangle,dashed, rounded corners, minimum width=2cm, minimum height=1cm,text centered, text width=2cm, draw=black, fill=red!30]
\tikzstyle{n} = [rectangle, rounded corners, minimum width=0cm, minimum height=0cm,note]

\usepackage{graphicx}
\graphicspath{ {../Figures/} }

\begin{document}

\title{Symbolic analysis meets federated learning to enhance malware identifier}

\author{Charles-Henry Bertrand Van Ouytsel }
\affiliation{%
  \institution{UCLouvain}
  \country{Belgium}}
\email{charles-henry.bertrand@uclouvain.be}

\author{Khanh Huu The Dam}
\affiliation{%
  \institution{UCLouvain}
  \country{Belgium}}
\email{khan.dam@uclouvain.be}
  
\author{Axel Legay}
\affiliation{%
  \institution{UCLouvain}
  \country{Belgium}}
\email{axel.legay@uclouvain.be}


\begin{abstract}
Over past years, the manually methods to create detection rules were no longer practical in the anti-malware product since the number of malware threats has been growing. Thus, the turn to the machine learning approaches is a promising way to make the malware recognition more efficient. The traditional centralized machine learning requires a large amount of data to train a model with excellent performance. To boost the malware detection, the training data might be on various kind of data sources such as data on host, network and cloud-based anti-malware components, or even, data from different enterprises. To avoid the expenses of data collection as well as the leakage of private data, we present a federated learning system to identify malwares through the behavioural graphs, i.e., system call dependency graphs.  It is based on a deep learning model including a graph autoencoder and a multi-classifier module. This model is trained by a secure learning protocol among clients to preserve  the private data against the inference attacks. Using the model to identify malwares, we achieve the accuracy of $\sim85\%$ for the homogeneous graph data and $\sim93\%$ for the inhomogeneous graph data.
\end{abstract}

%

\keywords{federated learning, symbolic analysis, malware detection, data privacy}


\maketitle

\section{Introduction}\label{sec:intro}

Since the number of malware threats has been growing over past years, the manually analysis methods to create detection rules were no longer practical in the anti-malware product. It thus requires the new advanced protection techniques which optimize the tasks of malware detection and classification. Hence, the turn to the machine learning (\ML) approaches is a promising way to make the malware recognition more efficient, robust and scalable~\cite{KasperLab, singh2021survey}. In the traditional centralized \ML, a large amount of data is required to train a model with excellent performance. To boost the malware detection, it might train the model on various kinds of data sources such as data on host, network and cloud-based anti-malware components, or even, data from different enterprises since the amount of data on each side is insufficient. Although all data can be collected in a central server for training, it is too expensive as well as it might cause the leakage of sensitive data~\cite{abadi2016deep,li2018learning}. 

To tackle this challenge, there are serveral works~\cite{lin2020using, 9014377, 2022108693, 9195012, Zhang2021, 10.1145/3368926.3369705, DBLP:conf/icdcs/NguyenMMFAS19} which implement the federated learning (\FL) which pushes model training to the devices from which data originate. Each device uses local data to train the local model, and then, all the local models are sent to the server to be aggregated into a global model. 
This \FL setting is risk in preserving data privacy since the sensitive information of the local data may be revealed via the inference attacks to training model~\cite{Wang2018, zhu2019deep, melis2019exploiting}. Hence, an additional privacy protection is setup to the \FL with amount of differential privacy noise adding to the model parameters~\cite{abadi2016deep}. However, the work in~\cite{boenisch2021curious} shows that relying on the differential privacy noise is insufficient to prevent the data leakage via the training model, as they explicitly trust the server with the crucial task of adding the differential privacy noise, and thus provide no protection against a semi-honest or untrusted server. In other hand, the more differential privacy noise is added to the model, the more degraded the performance of \FL.

Taking into account these issues, we introduce a \FL system that enables multiple participants to jointly learn a accurate deep learning model for malware detection and classification without sharing their input datasets while preserving privacy of the local data against the inference attacks. Our system consists of clients which hold their own datasets, a key-client which holds the secret key to encrypt the model parameters, and an aggregator which combines the local model parameters into a global model. To avoid the inference data attack, we replace the key-client for every training round. Even the aggregator or clients take snapshot of the training model, it is hard for them to infer the data from others. In particular, each client holds its own classifier model which is trained to recognize malwares at the clients' side. For this goal, we first characterise malwares by the system call dependency graphs (\scdg) which represent the program behaviours. Following ~\cite{SEBASTIO2020101775, bertrand2021detection,9277493}, we implement the symbolic analysis to explore all possible execution paths of a malware, and then, its corresponding \scdg is built from these execution paths.
Second, we propose a deep learning model including a graph autoencoder and a multi-classifier, to encode and classify the \scdgs. The graph autoencoder vectorizes the input \scdgs, and then, the multi-classifier identifies malwares according to the output of the autoencoder. We evaluate the deep learning model on two datasets. The first dataset includes \scdgs computed by the same extraction strategy. The second dataset includes \scdgs computed by different strategies.
In the experiment, we obtain significant results which are comparable to the \emph{Centralized Learning}. The accuracy is $\sim85\%$ for the homogeneous graphs. It can achieve $\sim93\%$ for the inhomogeneous graphs.

Section~\ref{sec:relatedwork} presents related works. Then, we recall the definition of system call dependency graph, and present the symbolic analysis to extract the graphs from malwares in Section~\ref{sec:scdg}. Section~\ref{sec:deepclassifier} presents the Deep neural classifier which we propose for encoding and classifying the system call dependency graphs. Our secure learning approach to train the deep learning model is introduced in Section~\ref{sec:FLconfiguration}. Section~\ref{sec:experiment} reports its evaluation results on our datasets. Finally, the conclusion is presented in Section~\ref{sec:conclusion}.

\section{Related work}\label{sec:relatedwork}
Malware detection approaches are mainly categorized into signature based and behavioural based approaches. The signature based approaches extract binary patterns for a malware family, and use these patterns to detect all samples of that family. The approaches suffer from building a huge database of signatures within a short period of time as well as they can not be used to detect new malware families. In addition, it is very easy to be overtaken by deforming or obfuscating the binaries. 
In the latter more advanced approaches, they detect malware by analyzing its behaviours instead of the syntax of the program. The behavioural analysis is usually distinguished as dynamic or static analysis. In the dynamic analysis \cite{egele2008survey}, malwares are executed in an emulated environment, e.g., sandbox, to look for a malicious behaviour~\cite{willems2007toward}. It is hard to trigger the malicious behaviour since there are limitation in execution time and the context emulated by the sandbox.

\noindent On the other hand, static techniques allow to analyze the behaviours of malware on the disassembled code without executing it. The static code analysis is performed concretely or symbolically. In the concrete analysis~\cite{gritti2020symbion}, the execution trace is computed from the disassembled code by some contextual information. Therefore, it has similar limitations to the dynamic analysis since the provided contexts cannot cover all the executions of program. 
The symbolic analysis performs a symbolic execution with symbolic input variables in place of concrete values. It allows exploring all the possible execution paths in the control flow graph~\cite{SEBASTIO2020101775,bertrand2021detection,9277493}. Thus, the malicious behaviours are easily exposed in this analysis. We consider the symbolic analysis in this work to construct \scdgs as behavioural signatures of malware. This behavioural graph representation has been proved as a good approach for malware detection in the recent works~\cite{7888729, said2018detection, bertrand2021detection, fredrikson2010synthesizing, lajevardi2021markhor, macedo2013mining, 10.1145/3412841.3441919}.

\medskip

\noindent \cite{7888729, bertrand2021detection} implement machine learning techniques on \scdgs for malicious behaviour extraction as well as malware detection. The graph mining techniques are employed to find malicious patterns in \scdgs of malwares in \cite{said2018detection, fredrikson2010synthesizing, macedo2013mining,lajevardi2021markhor}. \cite{10.1145/3412841.3441919} identifies malware families by applying the clustering algorithm on \scdgs. Those works obtain good results in the malware detection and classification. However, there are several challenges in training and updating the malware classifiers since malware detectors need to be updated consistently with a mass number of malwares. In addition, collecting all data in a centralized manner are so expensive while the data on individual devices are insufficient for training the efficient machine learning model.
Therefore, the federated learning is proposed to handle the issues in the traditional machine learning~\cite{konevcny2016federated}. It requires each device to use local data to train the local model, and then all the local models are uploaded to the server to be aggregated into a global model. This learning techinque is successfully applied for anomaly detection~\cite{DBLP:conf/icdcs/NguyenMMFAS19, rey2022federated, ghimire2022recent} as well as malware detection \cite{lin2020using, 9194132,galvez2020less}.
\cite{DBLP:conf/icdcs/NguyenMMFAS19, rey2022federated, ghimire2022recent} successfully employ the federated learning techniques for anomaly detection  on IoT devices. \cite{lin2020using} also shows the positive efficiency of the federated learning for malware classification comparing to the traditional machine learning, i.e., support vector machine (SVM). However, it is lack of the private data protection. Moreover, the privacy-preserving federated learning system is implemented in \cite{9194132,galvez2020less}. \cite{9194132} allows mobile devices to collaborate together for training a SVM classifier with a secure multi-party computation technique. \cite{galvez2020less} implements an average weight model to preserving the data privacy. Enhancing the models in \cite{galvez2020less,9194132}, we implement a secure learning approach with the homomorphic encryption to avoid the data inference attacks. Additionally, we implement a federated learning on inhomogeneous \scdgs to exploit the computation power of devices as well as the various representations of malware behaviours.

\section{System call dependency graph}\label{sec:scdg}

System call dependency graph (\scdg) is a directed graph consisting of nodes and edges, that represents the beahviours of a program. The nodes are (system) function calls. An edge represents information flowing between two two system calls in execution traces. The execution traces are obtained through the symbolic execution of a binary. Each trace is a list of system calls and the relevant information of these calls, such as the arguments, the resolved address and the calling address. Two system calls on an execution trace are data dependent if they are either argument relationship or address relationship. Two system calls have the argument relationship if they are using the same argument in an execution trace. The address relationship of two system calls is when an argument of a system call is the return value of another call.

 Let $S$ be a set of system calls. Formally,  a system call dependency graph is a directed graph  $G =(V,E,L)$ such that: $V$ is a set of nodes, $L: V \rightarrow S$ is a labelling function which maps a node $v \in V$ to a system call $s \in S$ and $E: V \times V$ is set of edges. $(v_1,v_2) \in E$ means that the system calls $L(v_1)$ and $L(v_2)$ are data dependent and the call to $L(v_1)$ is made before the call to $L(v_2)$,
e.g., the argument relationship of two system calls $\texttt{GetModuleFilenameA(0,m)}$ and $\texttt{CopyFileA(m,m',1)}$ is represented by an edge $(v_1,v_2) \in E$ in the graph $G=(V,E)$ such as $v_1, v_2 \in V$, $L(v_1) = \texttt{GetModuleFilenameA}$ and $L(v_2) = \texttt{CopyFileA}$.

\noindent In this work, we implement a symbolic execution to extracte \scdgs from the binaries as follows. First, the symbolic execution is performed on malware binaries to explore all possible execution paths. Thanks to \verb|angr| engine~\cite{7546500}, the execution flows of the binary are recorded as states including the instruction addresses, register values, memory usages, etc. in a period of time. At each execution step, if a variable can take several values, the symbolic value is replaced to keep track the execution and the related constraints. During the execution, the new states are created according to the instruction by the \verb|angr| engine. If the branch instruction, i.e., the conditional jump, the current state is forked into two states: the first is considered for taking the jump instruction, and the second is  considered for the next instruction. Otherwise, the current state produces a single child state. Hence, the exploration will produce a huge number of states during the symbolic execution. The state explosion can be controlled by using some constraint solvers, i.e., z3 or SMT solver, or heuristics to optimize the symbolic execution~\cite{SEBASTIO2020101775}. Following~\cite{bertrand2021detection}, we consider three strategies to explore the state at each execution step to compute \scdgs:
\begin{enumerate}
\item Custom  Breadth-first  search (\gone) implements the Breadth-first search to construct the graph of possible shortest paths from the execution traces.
\item Custom  Depth-first  search (\gtwo) considers the possible longest paths in the Depth-first  search on the execution traces.
\item Breadth-first search (\gthree) explores all paths from the execution traces in the Breadth-first search manner.
\end{enumerate} 
After the state exploration, we obtain serveral execution traces from a binary. Each trace is a sequence of system calls with the arguments, the resolved addresses and the calling addresses. Then, a \scdg is built on these execution traces. Its nodes correspond to system calls. Its edges represent the information flow of pair of system calls and their order in the execution traces. An edge is built from two system calls if they are either the argument relationship or the address relationship in the same execution trace. By using different exploration strategies, we may obtain various \scdgs to characterise the same binary. Hence, the study on such various \scdgs may boost the performance of malware detection as well as malware family recognition.

\section{Deep learning model}\label{sec:deepclassifier}
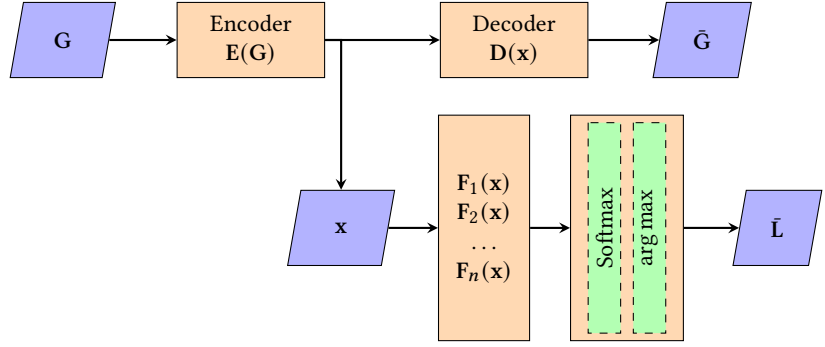
\begin{figure*}[!ht]
 \scalebox{1.0}{
\begin{tikzpicture}[node distance=1.5cm]
\node (input) [io] {$\mathbf{G}$};
\node (encoder) [process, right of=input,xshift=1cm] {Encoder $\mathbf{E}(\mathbf{G})$};
\node (decoder) [process, right of=encoder,xshift=2cm] {Decoder $\mathbf{D}(\mathbf{x})$};
\node (out1) [io, right of=decoder,xshift=1cm] {$\mathbf{\bar{G}}$};
\node (out2) [io, below of=encoder,xshift=1.2cm, yshift=-1cm] {$\mathbf{x}$};
\node (classifier) [classifier,right of=out2,xshift=0.4cm] {$\mathbf{F}_1(\mathbf{x})$ $\mathbf{F}_2(\mathbf{x})$ \\ $\ldots$ \\ $\mathbf{F}_n(\mathbf{x})$};
\node (logistic) [container,right of=classifier,xshift=0.4cm]{};
\node (softmax) [classifier90,right of=classifier,yshift=-1.6cm,xshift=-1.5cm] {Softmax};
\node (argmax) [classifier90,right of=softmax,yshift=-0.6cm,xshift=-1.5cm] {$\argmax$};
\node (out3) [io, right of=logistic,xshift=0.5cm] {$\mathbf{\bar{L}}$};

\draw [arrow] (input) -- (encoder);
\draw [arrow] (encoder) -- (decoder);
\draw [arrow] (decoder) -- (out1);
\draw [arrow] (encoder) -| (out2);
\draw [arrow] (out2) -- (classifier) ;
\draw [arrow] (classifier) -- (logistic);
\draw [arrow] (logistic) -- (out3);
\end{tikzpicture}
}
\caption{Deep neural network classifier}
 \label{fig:dnnc}
\end{figure*}

Since the \scdgs correspond to the behaviours of malware, we construct a deep learning model in order to encoding the \scdgs and classifying the malware.
The model consists in a graph autoencoder connecting to a classifier module, shown in Figure~\ref{fig:dnnc}. First, the graph autoencoder transforms a \scdg into a feature vector. Then, the classifier module takes the feature vector as input, and produces a predicted class of \scdg associated with this feature vector. They are presented in the following sections.
\subsection{Graph autoencoder}
\label{sec:graphencoder}
The graph autoencoder~\cite{10.1145/3412841.3441919} is equipped with Long Short Term Memory (\lstm) layers to embedding a \scdg into a feature vector.
A \lstm layer is a recurrent neural network  which is able to remember information for a long periods of times through memory cells. Each cell has three connected gates to control the internal state: the input gate $i_t$ is used to decide which part of information is stored in the memory cell; the forget gate $f_t$ is used to decide which part of information is thrown away from the memory cell; and the output gate $o_t$ which specifies the output.  Given an input $x_t$ and the output of previous cell $h_{t-1}$, the output $h_t$ is computed by $h_t = o_t \ast \textbf{tanh}(C_t)$. $C_t$ is the internal state of the current cell is computed as follows:
$$C_t = f_t \ast C_{t-1} + i_t \ast \tilde{C}_t$$  
where $\tilde{C}_t = \textbf{tanh}(W_C \cdot [h_{t-1},x_t]+b_C)$ is a new candidate for the internal state. 
Over periods of time, three gates of the memory cell are interacting layers with the sigmoid function $\sigma(\cdot)$ as follows:
$$f_t = \sigma(W_f \cdot [h_{t-1},x_t] +b_f)$$
$$i_t = \sigma(W_i \cdot [h_{t-1},x_t] +b_i)$$
$$o_t = \sigma(W_o \cdot [h_{t-1},x_t] + b_o)$$
where $\{W_f, W_i, W_o\}$ and $\{b_f,b_i,b_o\}$ are weights and biases of corresponding layers. These $\{W_f, W_i, W_o\}$ and $\{b_f,b_i,b_o\}$ are also called model parameters.

Let $\Gr$ be a \scdg. The graph autoencoder trains its \lstm layers through two modules: The encoder module $\EG$ transforms the graph structure, i.e., possible paths in $\mathbf{G}$, into a feature vector, i.e.,$\x= \EG$. The decoder module $\Dx$ does an opposite way to reconstruct the original graph $\Gbar$ from its feature vector $\x$ given by the encoder module, i.e., $\Gbar = \Dx$. This graph autoencoder is trained to optimize a reconstruction error $\text{Loss}_1$: 
$$\text{Loss}_1(\Gr,\Gbar)= \frac{1}{2} \sum_{i=1}^n ||c_i-\hat{c}_i||^2$$ 
where $||c_i-\hat{c}_i||$ is the differency measure of the component $c_i$ of the original graph $\Gr$ and its reconstructed component $\hat{c}_i$ in the graph $\Gbar$.  Using \lstm layers as its internal layers enables the graph autoencoder featuring  arbitrary size graphs, i.e., \scdgs. 

\subsection{Classifier module}
\label{sec:classifying}
The classifier module is a combination of $n$ single classifiers, i.e., $\{\mathbf{F}_i(\mathbf{x})\}^n_{i=1}$, $n> 0$. Each classiifer $i$ corresponds to a malware family. It is a fully connected neural network layer connected to the activation function $\text{softmax}(\cdot)$~\cite{Goodfellow-et-al-2016}. These classifiers are connected to the graph autoencoder via the output of the encoder $\EG$. Let $\x$ be an output of $\EG$. The module is defined as follows:

$$\hat{y} = \text{softmax}([\mathbf{F}_i(\mathbf{x})]^n_{i=1})$$
$$\Label(\x)  = \argmax^n_{i=1}(\hat{y})$$
where $\Label(\x)$ is the predicted label class of \scdg indicating the predicted malware family, $\mathbf{F}_i(\mathbf{x}) = \mathbf{W}_i\cdot \x + \mathbf{b}_i$. $\mathbf{W}_i$ and $\mathbf{b}_i$ respectively are  weight and bias values of the layer $\mathbf{F}_i$. The $\{(\mathbf{W}_i, \mathbf{b}_i)\}_{i=1}^n$ are model's parameters. 
\\
This module is trained to optimize the Binary Cross Entroy Loss: 
$$\text{Loss}_2 = -\frac{1}{n} \sum_{i=1}^n y_i \log(\hat{y}_i) + (1-y_i )\log(1-\hat{y}_i)$$
where $y$ specifies the ground truth class of \scdg associated with $\x$. $y_i=1$ if the \scdg is in the class $i$. Otherwise, $y_i=0$.

\medskip

\noindent To train the deep learning model, we use a stochastic optimization algorithm, i.e., Adam optimization algorithm~\cite{kingma2014adam}, which recomputes the model parameters in very training step on the training data, in order to optimizing the total loss function. The total loss function $\text{Loss}$ is a sum of losses from the graph autoencoder and the classifier module. 
$$\text{Loss} = \text{Loss}_1  + \text{Loss}_2.$$

\section{Secure learning approach}\label{sec:FLconfiguration}
In the previous section, we introduce the deep learning model to classify \scdgs. We present in this section a secure learning approach to train this model on different devices/clients using the concept of federated learning. We first recall the federated learning (\FL)~\cite{konevcny2016federated}. \FL is a collaboratively learning among \textbf{N} devices/clients with the help of a central server, e.g., an aggregator server. In the training process, clients use their private dataset to train their local model, and then, all the local models are sent to the aggregator server to be aggregated for a global model. Then, the parameters of the global model are sent to clients for updating to their local model.
After a sufficient number of local training and updates exchanges between the aggregator server and the associated clients, the clients' models can converge to an optimal learning model. However, the \FL is not always a sufficient privacy mechanism to protect the privacy of clients, i.e., the private dataset. It is easy to be exposed to the data inference attacks. As we have mentioned above, the popular technique, such as using differential privacy to protect the private data, is not strong enough to protect the leakage of private data \cite{boenisch2021curious}. In addition, using encryption algorithms, such as using homomorphic encryption \cite{8241854, 10.1145/2857705.2857731}, multi-party computating (MPC) \cite{10.5555/1929820.1929840}, is more potential ways to preserve the data privacy. 
In the section, we implement a secure learning protocol with an average-weight aggregation using homomorphic encryption. It protects the private data from semi-honest clients and server. Note that the semi-honest (a.k.a honest-but-curious) model is that collaborators (clients/server) do follow the protocol but try to infer as much as possible from the values (shares) they learn, also by combining their information. This protocol is presented in the following sections.

\subsection{Communication setting}
\label{sec:comm}
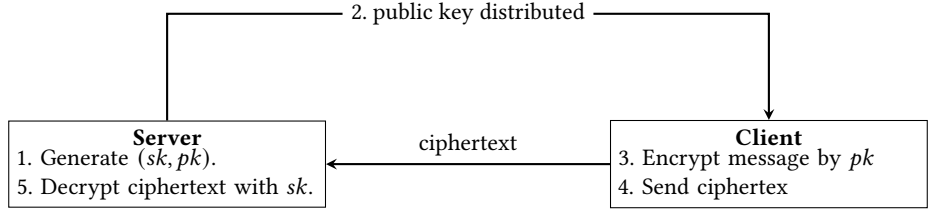
\begin{figure*}[!ht]
 \scalebox{1.0}{
\begin{tikzpicture}[node distance=2cm]

\node (server) [rectangle, minimum width=1cm, minimum height=1cm, text centered,text width=4cm, draw=black] {\textbf{Server} \\ \parbox{\textwidth}{1. Generate $(sk,pk)$. \\ 5. Decrypt ciphertext with $sk$.}};

\node (client) [rectangle, minimum width=1cm, minimum height=1cm, text centered,text width=4cm, draw=black, right of= server,xshift=6cm] {\textbf{Client} \\ \parbox{\textwidth}{3. Encrypt message by $pk$ \\ 4. Send ciphertex}};
\node (n1) [rectangle, above of=server,right of=server,xshift=2cm] {2. public key distributed};

\draw [arrow] (server) |- (n1) -| (client);
\draw [arrow] (client) --node[anchor=south]{ciphertext} (server);
\end{tikzpicture}
}
\caption{Client-Server communication}
\label{fig:communication}
\end{figure*}

We implement an asymmetric encryption for the communication between clients and the server (Figure~\ref{fig:communication}). Support a client want to send safely a message to the server. First, the server generates a key pair composed of a private key and a public key. The private key is hidden by the server. The public key is openly distributed to the client. The private key can decrypt what the public key encrypts, and vice versa. Then, the client encrypts its message by using the server's public key. It openly sends this ciphertext to the server. This ciphertext is safe since any other cannot decrypts it without the private key from the server. Server receives the ciphertext and decrypts it with its private key.
Thus, this setting ensures a secure communication between the client and the server.

\subsection{Homomorphic encryption for a secure training model}
For a safe computation at the aggregator, we implement the homorphic encryption to the model parameters.
let us recall the homomorphic encryption scheme~\cite{regev2009lattices, chowdhary2021eva, kim2020approximate}. Let $(\mathbf{sk},\mathbf{pk})$ be a pair of secret and public keys, respectively. Given a numeric value $\mathbf{z}$. First, $\mathbf{z}$ is encoded into a plaintext. The public key $\mathbf{pk}$ is used to encrypt the plaintext of $\mathbf{z}_{plaintext}$ into a ciphertext $\bar{z}_{\mathbf{pk}}$. 
Let $\text{HEnc}$ be a function which encodes a numeric value and encrypts its plaintext into a ciphertext with a public key, e.g., $\bar{z}_{\mathbf{pk}} = \text{HEnc}(\mathbf{z},\mathbf{pk})$. Let $\text{HDecrypt}$ be a function which decrypts a ciphertext by using a secret key and decodes its plaintext into a numeric value.
Then, $\text{HDecrypt}(\bar{z}_{\mathbf{pk}},\mathbf{sk})$ decrypts $\bar{z}_{\mathbf{pk}}$ to the plaintext $\mathbf{z}_{plaintext}$, and it decodes $\mathbf{z}_{plaintext}$ to the numeric value $\mathbf{z}$, i.e., $\mathbf{z}= \text{HDecrypt}(\bar{z}_{\mathbf{pk}},\mathbf{sk})$. 

The important property of homomorphic encryption is that users can process and calculate the encrypted data without revealing the original data. With the secret key, the user decrypts the processed data, which is exactly the expected result~\cite{chen2018privacy, nikolaenko2013privacy}.
Given numeric values $\mathbf{x}$ and $\mathbf{z}$, the multiplication of the ciphertext of $\mathbf{z}$, i.e., $\bar{z}_{\mathbf{pk}}$, and the plaintext of $\mathbf{x}$, i.e., $\mathbf{x}_{plaintext}$, is a ciphertext of $\mathbf{z} \cdot \mathbf{x}$, i.e., $\text{HEnc}(\mathbf{z}\cdot\mathbf{x},\mathbf{pk}) = \bar{z}_{\mathbf{pk}} \cdot \mathbf{x}_{plaintext}$. 
The addtion of two ciphertexts $\bar{x}_{\mathbf{pk}}$ and $\bar{z}_{\mathbf{pk}}$ is also a ciphertext of $\mathbf{z}+\mathbf{x}$, i.e.,  $\text{HEnc}(\mathbf{z}+\mathbf{x},\mathbf{pk}) = \bar{z}_{\mathbf{pk}} + \bar{x}_{\mathbf{pk}}$.

\medskip

\noindent With the homomorphic encryption, we implement a secure learning among \textbf{N} clients. The local updates is protected in the encrypted message. The global update is computing on encrypted data with respect to the addition operator of the homomorphic encryption. The Algorithm \ref{alg1} describes the $m$ training rounds of \textbf{N} clients. Each client $i$ holds a pair of $(\mathbf{pk}^i, \mathbf{sk}^i)$, and the \emph{Key-client} holds a secret value $\mathbf{z}$. For each training round, the training process is as follows: First, the system selects a \emph{Key-client} among clients. The selected client, i.e., \emph{Key-client}, exchange the public key, i.e., $\mathbf{pk}^i$, and its encrypted value, i.e., $\bar{z}_{\mathbf{pk}^s} = \text{HEnc}(\mathbf{z}, \mathbf{pk}^s)$ with other clients. Meanwhile, each client locally trains their local model. When the training step is done, they use the ciphertext $\bar{z}_{\mathbf{pk}^s}$ of the \emph{Key-client} to encrypt the parameters of their local model and send to the \emph{aggregator}. Then, the \emph{aggregator} computes the sum of clients's parameters, and sends to the \emph{Key-client}. The \emph{Key-client} decrypts the global updates from the \emph{aggregator}, i.e., $\mathbf{w} = \frac{1}{\mathbf{z}\cdot\mathbf{N}}\text{Decrypt}(\bar{w}_{\mathbf{pk}^s})$ and sends the updates $\mathbf{w}$ to every client $i$. Finally, the clients receive and apply the updates to their local model. After these updates are done, a new round is started by voting a new \emph{Key-client} and training the local models at the clients' side.

\begin{algorithm} 
\begin{algorithmic}[1]
\ForEach{$ round \in [1 \ldots m]$ }
\State Randomly vote a \emph{Key-client} among clients.
\State The clients send their public keys, i.e., $\mathbf{pk}^i$, to the \emph{Key-client}.
\State The \emph{Key-client} sends its encrypted secret value $\bar{z}_{\mathbf{pk}^s} = \text{HEnc}(\mathbf{z},\mathbf{pk}^s)$ to all clients.
\ForEachpara{$ i \in [1 \ldots \mathbf{N}]$ }
\State The client $i$ trains the local model on its own data.
\State The client $i$ encrypts the model's parameters $\mathbf{w}^i$ using the secret value $\bar{z}_{\mathbf{pk}^s}$, i.e., $\bar{w}^i_{\mathbf{pk}^s}= \mathbf{w}^i\cdot \bar{z}_{\mathbf{pk}^s}$.
\State The client $i$ sends the update $\bar{w}^i_{\mathbf{pk}^s}$ to the \emph{aggregator}.
\EndFor
\State The \emph{aggregator} computes the encrypted global weight $$\bar{w}_{\mathbf{pk}^s}= \sum_{i=1}^n \bar{w}^i_{\mathbf{pk}^s}.$$
\State The \emph{aggregator} sends the encrypted global weight $\bar{w}_{\mathbf{pk}^s}$ to the \emph{Key-client}.
\State The \emph{Key-client} decrypts $\bar{w}_{\mathbf{pk}^s}$ by its secret key $\mathbf{sk}^{s}$ and the secret value $\mathbf{z}$ to get the global weight $\mathbf{w}$.
$$ \mathbf{w} = \frac{1}{\mathbf{z}\cdot \mathbf{N}}\text{HDecrypt}(\bar{w}_{\mathbf{pk}^s},\mathbf{sk}^{s})$$
\ForEachpara{$ i \in [1 \ldots \mathbf{N}]$ }
\State The \emph{Key-client} encrypts the updates $\mathbf{w}$ by $\mathbf{pk}^i$, i.e., $\mathbf{w}_{\mathbf{pk^i}}$,  and sends the updates to the client $i$.
\State The client $i$ decrypts $\mathbf{w}_{\mathbf{pk^i}}$ by its secret key $\mathbf{sk}^i$ and applies the updates to its local model.
\EndFor
\EndFor
\end{algorithmic}
\caption{Secure training in the federated learning}
\label{alg1}
\end{algorithm}

\subsection{Application to malware identifier}
\label{sec:app_mal}
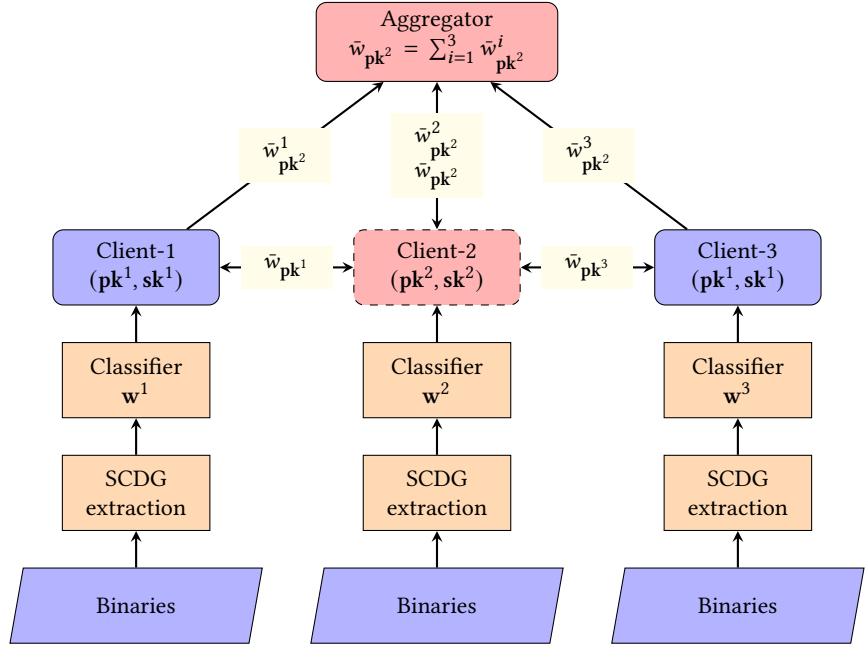
\begin{figure*}[!ht]
 \scalebox{1.0}{
\begin{tikzpicture}[node distance=2cm]

\node (aggregator) [server] {Aggregator \\ $\bar{w}_{\mathbf{pk}^2} =\sum_{i=1}^3 \bar{w}^i_{\mathbf{pk}^2}$};

\node (client2) [client1,below of=aggregator,yshift=-1cm] {Client-2 \\ $(\mathbf{pk}^{2},\mathbf{sk}^{2})$};
\node (idx2) [process,below of=client2,yshift=0.5cm] {Classifier \\ $\mathbf{w}^2$};
\node (gex2) [process,below of=idx2,yshift=0.5cm] {\scdg extraction};
\node (inp2) [io,below of=gex2,yshift=0.5cm] {Binaries};
\draw [arrow] (inp2)-- (gex2);
\draw [arrow] (gex2)-- (idx2);
\draw [arrow] (idx2)-- (client2);

\node (client3) [client0, right of=client2,xshift=2cm] {Client-3 \\ $(\mathbf{pk}^{1},\mathbf{sk}^{1})$};
\node (idx3) [process,below of=client3,yshift=0.5cm] {Classifier \\ $\mathbf{w}^3$};
\node (gex3) [process,below of=idx3,yshift=0.5cm] {\scdg extraction};
\node (inp3) [io,below of=gex3,yshift=0.5cm] {Binaries};
\draw [arrow] (inp3)-- (gex3);
\draw [arrow] (gex3)-- (idx3);
\draw [arrow] (idx3)-- (client3);

\node (client1) [client0, left of=client2,xshift=-2cm] {Client-1 \\ $(\mathbf{pk}^{1},\mathbf{sk}^{1})$};
\node (idx1) [process,below of=client1,yshift=0.5cm] {Classifier \\ $\mathbf{w}^1$};
\node (gex1) [process,below of=idx1,yshift=0.5cm] {\scdg extraction};
\node (inp1) [io,below of=gex1,yshift=0.5cm] {Binaries};
\draw [arrow] (inp1)-- (gex1);
\draw [arrow] (gex1)-- (idx1);
\draw [arrow] (idx1)-- (client1);

\draw [arrow0] (client1) --node[note] { $\bar{w}_{\mathbf{pk}^1}$} (client2);
\draw [arrow0] (client3) --node[note] { $\bar{w}_{\mathbf{pk}^3}$} (client2);

\draw [arrow] (client1) --node[note] { $\bar{w}^1_{\mathbf{pk}^2}$} (aggregator);
\draw [arrow0] (client2) --node[note] { $\bar{w}^2_{\mathbf{pk}^2}$ \\ $\bar{w}_{\mathbf{pk}^2}$} (aggregator);
\draw [arrow] (client3) --node[note] {$\bar{w}^3_{\mathbf{pk}^2}$} (aggregator);

\end{tikzpicture}
}
\caption{Secure learning among $3$ clients and one aggregator. Client-2 is the \emph{Key-Cient}, and it might be replaced by another after each training round.}
\label{fig:ex_secure_training}
\end{figure*}

We apply the secure learning protocol to train the deep learning model on $N$ clients. Each client has its own malwares which cannot be shared to others. To identify malwares, the clients implement a classifier model at their side. They share the parameters of their local classifier model during the training phase. They get back the update from the \emph{aggregator} after each training round. An example of the training process of 3 clients is shown in Figure~\ref{fig:ex_secure_training}: Clients keep their binaries/malwares. Using the \scdg extraction, they generate \scdgs from their binaries, and the \scdgs are also kept in private. Then, they train the local model on the extracted \scdgs. The trained model's parameters are encrypted,  and they are sent to \emph{aggregator}. The aggregator safely computes the update on the encrypted parameters. Then, it sends the aggregated parameters to the \emph{Key-client}, i.e., Client-2, for decryption. The \emph{Key-client} decrypts the aggregated parameters, and sends them to Client-1 and Client-3. The clients receive the aggregated parameters, and update to the local model. The communication channels among the aggregator and clients are implemented following the client-server communication in Section~\ref{sec:comm}.

\medskip

\noindent In this work, we consider two cases of sharing the model paramaters as follows:
\begin{enumerate}
\item Clients collaboratively learn the full model including the graph autoencoder $(\EG,\Dx)$ and the classifier module $\{\mathbf{F}_i(\cdot)\}$, called \emph{Full-Aggregation Learning}. This is a closed collaboration since all clients should have the same structure of their local model as well as they share their data labels.

\medskip

\item We decompose the model in Section \ref{sec:deepclassifier} into two parts: the autoencoder $(\EG,\Dx)$ and the classifier module $\{\mathbf{F}_i(\cdot)\}$. Clients share only a part of model, i.e., the encoder part $\EG$, with each other, called \emph{Partly-Aggregation Learning}. It enables clients sharing their feature computing in the form of an encoder $\EG$. Thus, it keeps the private data classes. It then allows a flexible implement of learning paradigms at the client's side.
\end{enumerate}

\section{Experiment}\label{sec:experiment}
We evaluate the performance of our deep learning system in this section. We first present our datasets and the data proportion on each client. Then, we present the implement of our learning approaches on the homogeneous \scdgs and on the inhomogeneous \scdgs. The results also are compared the implement of the \emph{Centralized Learning}. 
The \FL experiment is deployed on 4 virtual machines. Their resources are reported in Table~\ref{tab:vm-resources}.
\begin{table}[!ht]
\begin{tabular}{|l|c|c|c|}
\hline
 & Physical CPU &	Virtual CPU &	Memory \\
 \hline
Aggregator & 4 &	8 &	24GB \\
\hline
Client-1 & 4 &	8 &	24GB \\
\hline
Client-2 & 8 &	8 &	20GB \\
\hline
Client-3 & 8 &	8 &	15GB \\
\hline
\end{tabular}
\caption{Virtual Machines used in our experiment}
\label{tab:vm-resources}
\end{table}
The evaluation is measured by the accuracy of the trained malware identifier as follows:
$$\text{Accuracy}(y,\hat{y}) = \frac{1}{n}\sum_{i=1}^n 1(y_i=\hat{y}_i)$$
where $n$ is the number of samples, $1(\cdot)$ is the indicator function, $\hat{y}_i$ is the predicted value of the i-th sample and $y_i$ is the corresponding true value.

\subsection{Dataset}
We collect malwares from Cisco and MalwareBazaar Database~\footnote{https://bazaar.abuse.ch/} to  build two datasets: \dsone and \dstwo. \dsone consists of 2260 malwares from 15 families. It is randomly splitted into two sets: the training set of 2034 malwares and the test set of 226 malwares. The \scdgs in \dsone are computed by the same strategy, i.e., \gtwo which is presented in Section~\ref{sec:scdg}. \dsone is used for the \emph{Homogeneous-data} scheme. \dstwo is a collection of 1844 malwares from 15 families. They are distributed to three clients. Each client implements its own strategy to compute \scdgs from binaries in \dstwo. Particularly, Client-1 successfully extracts 1660 graphs by strategy \gthree. Using strategy \gone, Client-2 successfully extracts 1725 graphs. Client-3 obtains 1662 graphs by strategy \gtwo. Since the clients implement different strategies, in a limit period of time, i.e., 20 minutes, the number of \scdgs extracted from \dstwo is different at each client. Thus, this challenges our secure learning model to deal with  the context of inhomogeneous \scdgs. The data proportion of each dataset are detailed in Table~\ref{tab:dataset}.
\begin{table}
\centering
\begin{tabular}{|c|c|c|c|}
\hline
 & Partitions & Training set & Test set \\
\hline
\hline
\multirow{3}{*}{\dsone} & Client-1 & 678 & \multirow{3}{*}{226} \\
\cline{2-3}
& Client-2& 678 & \\
\cline{2-3}
& Client-3& 678 & \\
\hline
\multicolumn{2}{|c|}{\bf Total} & 2034 & 226 \\
\hline
\hline
\multirow{3}{*}{\dstwo} & Client-1 & 1245 &  415\\
\cline{2-4}
& Client-2& 1293 & 432\\
\cline{2-4}
& Client-3& 1246 & 416\\
\hline
\end{tabular}
\caption{The distribution of data in \FL}\label{tab:dataset}
\end{table}

\subsection{Secure Federated Learning on \scdgs}
In this experiment, our approach is evaluate on a homogeneous dataset, i.e., \dsone. First, we evaluate the model on the centralized data of \dsone. Then, we compare this result to the implement of our secure learning approach  where the data of \dsone are splitted into 3 partitions of 678 malwares, and they are distributed to 3 clients. The data proportion at clients is shown in Table~\ref{tab:dataset}. 

\paragraph{Centralized Learning:}  We setup the classifer to take a \scdg as input, and it outputs the feature vector of size of 64. The classifier module is constructed according to number of malware families in the training partition, i.e., 15 classifiers which correspond to 15 malware families. We train the model for 10 epoches on the training set. Then, the trained model is used to identify malwares from the test set. We obtain the accuracy of 85.4\%.

\paragraph{Secure Federated Learning:}
We implement the secure training for 3 clients, using the library TenSEAL~\cite{tenseal2021} for encrypting and aggregating the model parameters, and RSA cryptosystem\footnote{https://cryptography.io/} for a secure communication. The local models are trained on the client's training set. The updated models are evaluated on the test set. Note that the client's training set is a part of the training set in the \emph{Centralized Learning}, and the test set is used for both the \emph{Centralized Learning} and the \emph{Secure Federated Learning}. The proportion of data at clients is shown in Table~\ref{tab:dataset}.

\medskip

\begin{figure}[!ht]
\centering
\begin{tabular}{c}
\includegraphics[width=0.45\textwidth]{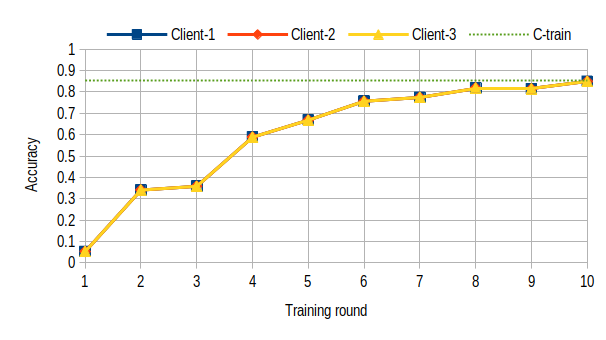}\\
(a) Full-aggregation learning\\
\includegraphics[width=0.45\textwidth]{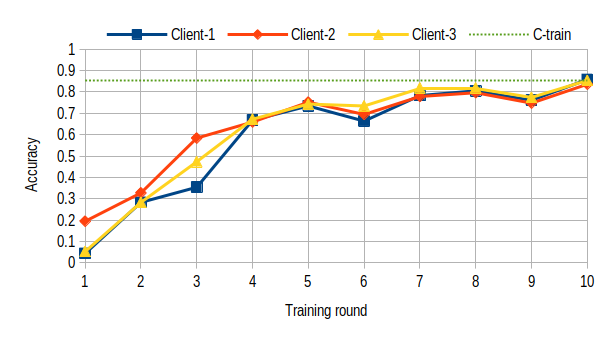}\\
(b) Partly-aggregation learning
\end{tabular}
\caption{Accuracy of the secure federated learing. The dash-line (C-train) is the accuracy of the centralized learning.}
\label{fig:acc1}
\end{figure}

\noindent Figure~\ref{fig:acc1} shows that the performance of the local models is improved after 10 training rounds. For the \emph{Full-Aggregation Learning} (Figure~\ref{fig:acc1}-a), the performance of all clients are similar, and their accuracy can reach 84.96\% comparing to the accuracy of 85.4\% in the \emph{Centralized Learning}. Although the increment of accuracy is a bit different among clients in the \emph{Partly-Aggregation Learning}, they are getting more converged at the end of the training phase. Comparing to the \emph{Centralized Learning}, the performance of Client-1 and Client-3 in \emph{Partly-Aggregation Learning} is equal to or even better than the ones in the \emph{Centralized Learning}. The results are also reported in Table~\ref{tab:summary1}.

\begin{table}[!ht]
\centering
\begin{tabular}{|c|c||c|}
\hline
\multicolumn{2}{|l||}{} & Accuracy ($\%$) \\
\hline 
\multicolumn{2}{|l||}{\emph{Centralized Learning}} & 85.4 \\
\hline
\hline 
\multicolumn{2}{|l||}{\emph{Full-Aggregation Learning}} & 84.96 \\
\hline
\hline 
\multirow{3}{*}{\emph{Partly-Aggregation Learning}} 
& Client-1 &\bf 85.84\\
\cline{2-3}
& Client-2& 83.63\\
\cline{2-3}
& Client-3&\bf 85.4\\
\hline
\end{tabular}
\caption{Comparison of the centralized learning and the secure federated learning.}
\label{tab:summary1}
\end{table}

\subsection{Secure Federated Learning on the inhomogeneous \scdgs}
Since the computing power is different from devices, the features, e.g., \scdg, are extracted accrordingly. The training graphs are computed in different techniques among clients even though that they are \scdgs. In this work, we consider three types of \scdg, that are extracted from binaries by the three different strategies in Section~\ref{sec:scdg}, i.e., \gthree, \gone and \gtwo, at three clients. Particularly, Client-1 implements \gthree, Client-2 implements \gone, and Client-3 implements \gtwo. The data proportion is reported in Table~\ref{tab:dataset}.
In the scheme, the clients have their own training set and test set. The training sets are used to train their models. Then, the test sets are used to evaluate the performance of the models. 
Similar to previous experiment, we first implement the \emph{Centralized Learning} in the client's side. Then, we compare this results to the secure learning approach with 2 aggregation cases.

\begin{enumerate}
\item For the \emph{Centralized Learning}, the model is separately trained on its training set for each client in 20 epoches. Then, this model is evaluated on the client's test set.

\medskip

\item For the \emph{Secure Federated Learning}, we implment two types  of aggregations: \emph{Full-Aggregation Learning} and \emph{Partly-Aggregation Learning} (see Section~\ref{sec:app_mal}). The secure training is applied to train the local model for 20 training rounds.  In each training round, the client trains the local model on its training set, and then, it evaluates the updated model on its own test set.
\end{enumerate}

\begin{figure}[!ht]
\centering
\begin{tabular}{c}
\includegraphics[width=0.45\textwidth]{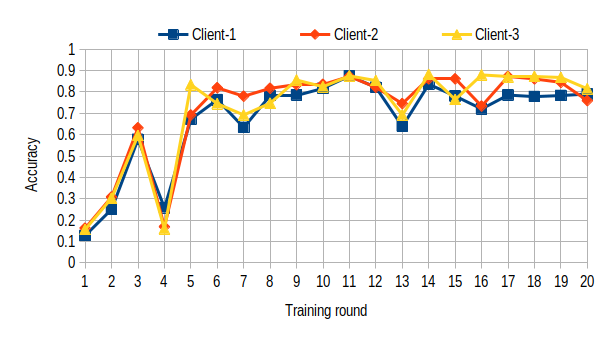}\\
(a) Full-aggregation learning\\
\includegraphics[width=0.45\textwidth]{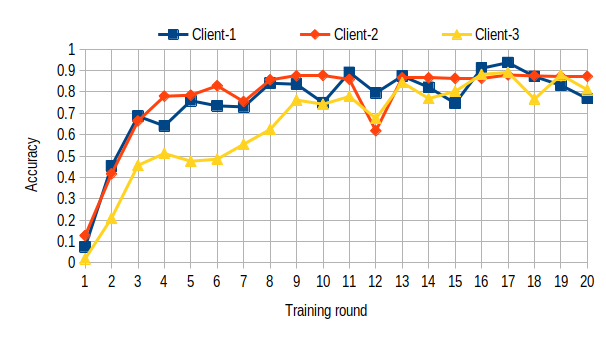}\\
(b) Partly-aggregation learning
\end{tabular}
\caption{Accuracy of the secure federated learing in the inhomogeneous-data scheme.}
\label{fig:acc2}
\end{figure}

\begin{table}[!ht]
\centering
\begin{tabular}{|c|c||c|}
\hline
\multicolumn{2}{|l||}{} & Accuracy ($\%$) \\
\hline 
\multirow{3}{*}{\emph{Centralized Learning}} 
& Client-1 & 92.05\\
\cline{2-3}
& Client-2& 81.02\\
\cline{2-3}
& Client-3& 89.9\\
\hline
\hline
\multirow{3}{*}{\emph{Full-Aggregation Learning}} 
& Client-1 & 87.4 \\ 
\cline{2-3}
& Client-2&\bf 87.27 \\ 
\cline{2-3}
& Client-3& 88.22 \\ 
\hline
\hline
\multirow{3}{*}{\emph{Partly-Aggregation Learning}} 
& Client-1 & \bf 93.73 \\ 
\cline{2-3}
& Client-2& \bf 87.96 \\ 
\cline{2-3}
& Client-3& 89.18 \\ 
\hline
\end{tabular}
\caption{Comparison of the centralized learning and the secure federated learning in the inhomogeneous-data scheme.}
\label{tab:summary2}
\end{table}

\medskip

\noindent Figure~\ref{fig:acc2}-a shows that the performance of the local models is getting improvement after 20 training rounds in \emph{Full-Aggregation Learning}. Three clients can reach the accuracy of $\sim 87\%$ after 11 training rounds. Then, the performance of Client-1 goes down to $79\%$ while Client-3 reach its peak, i.e., $88.22\%$, at the 14-th round. Compare to the \emph{Centralized Learning}, Client-2 in \emph{Full-Aggregation Learning} can achieve the better performance at the 8-th round. It reach 87.27\% of accuracy, at the 11-th round while the accuracy in the \emph{Centralized Learning} is 81.02\%. The degradation at Client-1 in \emph{Full-Aggregation Learning} is about $5.1\%$ comparing to the \emph{Centralized Learning}.
For \emph{Partly-Aggregation Learning} in Figure~\ref{fig:acc2}-b,  Client-1 gets the accuracy of 93.73\% after 17 training rounds while Client-2 and Client-3 get 87.96\% and 89.18\%, respectively. The performance of Client-1 and Client-2 overtake the ones in \emph{Centralized Learning} while the degradation at Client-3 is about 0.72\%. The results are also reported in Table~\ref{tab:summary2}.

\medskip

\noindent Overall, our secure learning approach is able to train the deep learning model for an accurate malware identifier. Comparing to the  \emph{Centralized Learning}, the performance gets a bit  degradation in some client. It is a side-effect of calculation on the encrypted data in the secure aggregation. Besides, the experimental results also show that the \emph{Partly-Aggregation Learning} overtakes the \emph{Full-Aggregation Learning} in our system. Hence, the \emph{Partly-Aggregation Learning} should be considered in the future implement of the federated learning for malware classification since it allows the clients collaborate with each other while keeping their classifier model in private.

\section{Conclusion}\label{sec:conclusion}
In this paper, we present a deep learning model for malware classification and a secure learning approach to integrate this learning model into a feretated learning system. We validate the system to identify malwares on our datasets. We obtains a significant result with the accuracy of $\sim 85\%$. It is comparable to the \emph{Centralized learning}. Moreover, we implement the system to learn \scdgs extracted from 3 various strategies. We can achieve the accuracy of $\sim88\%$ in \emph{Full-Aggregation Learning} and $\sim93\%$ in \emph{Partly-Aggregation Learning}, comparing to the accuracy of $\sim92\%$ in \emph{Centralized Learning}. According to the experimental results, the \emph{Partly-Aggregation Learning} is a good option to employ a collaboratively learning for malware classification in future.


\bibliographystyle{ACM-Reference-Format}
\bibliography{refs}


\end{document}